\documentclass{article}

    \usepackage[preprint,nonatbib]{neurips_2021}

\usepackage[utf8]{inputenc} 
\usepackage[T1]{fontenc}    
\usepackage{hyperref}       
\usepackage{url}            
\usepackage{booktabs}       
\usepackage{amsfonts}       
\usepackage{nicefrac}       
\usepackage{microtype}      
\usepackage{xcolor}         
\usepackage{authblk}
\newcommand{\etal}{\textit{ et al}.}

\usepackage{amssymb}
\usepackage{amsthm}
\usepackage{mathtools}
\usepackage{wrapfig}
\usepackage{caption}
\usepackage{subcaption}
\title{Deep Bayesian Learning for Car Hacking Detection}

\author[1]{Laha Ale}
\author[1]{Scott A. King}
\author[2]{Ning Zhang}

\affil[1]{Department of Computing Sciences, Texas A\&M University-Corpus Christi} 
\affil[2]{Department of Electrical and Computer Engineering, University of Windsor}
\affil[ ]{\textit {\{lale.islander, scott.king\}@tamucc.edu, ning.zhang@uwindsor.ca}}
\affil[ ]{\textit {{\dag \color{blue}https://github.com/ainilaha/ppl\_car\_hacking}}}

\begin{document}

\maketitle

\begin{abstract}
With the rise of self-drive cars and connected-vehicles, cars are equipped with various devices to assistant the drivers or support self-drive systems. Undoubtedly, cars have become more intelligent as we can deploy more and more devices and software on the cars. Accordingly, the security of assistant and self-drive systems in the cars becomes a life threatening issue as smart cars can be invaded by malicious attacks that cause traffic accidents. Currently, canonical machine learning and deep learning methods are extensively employed in car hacking detection. However, machine learning and deep learning methods can easily be overconfident and defeated by carefully designed adversarial examples. Moreover, those methods cannot provide explanations for security engineers for further analysis. In this work, we investigated Deep Bayesian Learning models to detect and analyze car hacking behaviors. The Bayesian learning methods can capture the uncertainty of the data and avoid overconfident issues. Moreover, the Bayesian models can provide more information to support the prediction results that can help security engineers further identify the attacks. We have compared our model with deep learning models and the results show the advantages of our proposed model. The code of this work is publicly available\dag.
\end{abstract}

\section{Introduction}
With the rise of self-driving cars and Vehicle to Everything (V2X), cars are connected to various devices through wireless internet. Cars are exposed to malicious attacks which may lead to severe traffic accidents. To mitigate these issues, various detection methods have be proposed to analyze and detect intrusions. However, the current existing methods can be overfitting and blinded by carefully designed attacks. Moreover, these methods have no good explanations for the prediction results that support security engineers on further analysis.

Controller Area Network (CAN) is a key component for a connected-vehicle and vulnerable to malicious attacks. Therefore, many researchers have attempted to find reliable intrusion detection methods to detect and analyze the attacks. Anomaly detection methods have been adopted to detect the attacks aimed at vehicles~\cite{Marchetti7995934,Muter5940552}. However, the anomaly detection can only distinguish normal operations and abnormal ones and without further information attached to the prediction results. Seo\etal~\cite{Song8514157} proposed an Intrusion Detection System (IDS) based on Generative Adversarial Networks (GAN)~\cite{Goodfellow2014}. More recently, Convolutional Neural Networks (CNN) have also been adopted to detect the attacks. Song\etal~\cite{SONG2020100198} proposed a heavy deep learning model based on Inception-ResNet~\cite{szegedy2016inceptionv4} to detect CAN intrusions, and Hossain\etal~\cite{Hossain9322395} used a simple CNN and achieved about the same performance. Although the existing methods can achieve considerably high accuracy in detecting the attacks, they cannot capture the uncertainty of the attacks. Moreover, those canonical machine learning and deep learning methods require a massive number of training examples with human effort to label the data set. 

In this work, we adopted Bayesian Deep Learning  to detect the CAN intrusion. With the proposed Bayesian method, we can provide uncertainty of the prediction and provide more information for security engineers. The Bayesian networks can also capture the uncertainty and capture adversarial attacks. The contributions of this work can be summarized as:
\begin{itemize}
 \item We developed a Deep Bayesian model for detecting attacks against vehicles that can determines uncertainty of its predictions.
 
 \item We introduce a Controller Area Network Intrusion Detection System (IDS) with Deep Bayesian and its training process.
 
 \item We present an analysis of the results of using Bayesian predictions on attack detection.
\end{itemize}

\section{Bayesian Networks for Intrusion Detection}


\begin{wrapfigure}{r}{0.5\textwidth}
    \centering
    \includegraphics[width=0.5\textwidth]{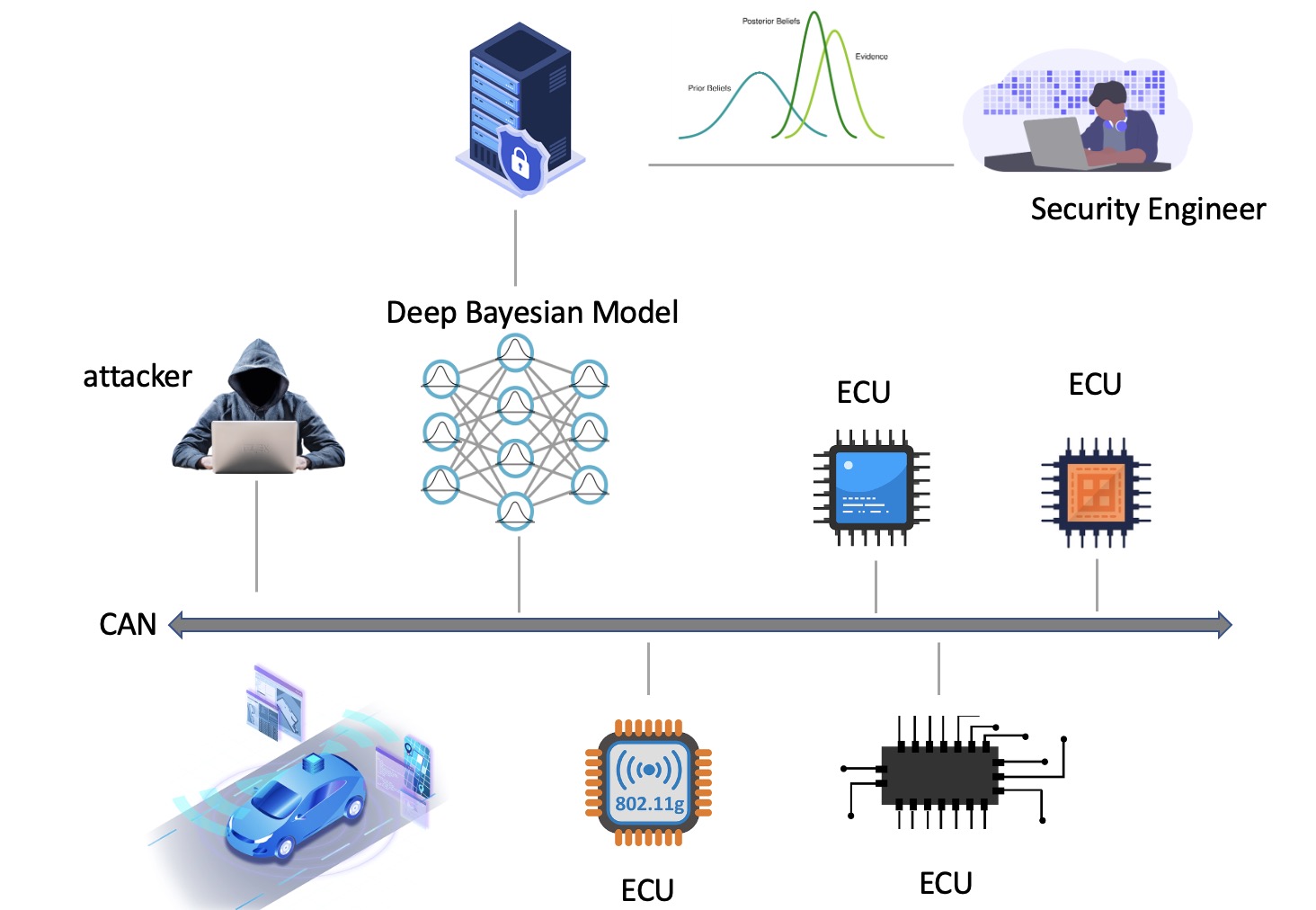}
    \caption{Detection and Training Process.}
    \label{fig:system}
\end{wrapfigure}
The Intrusion Detection System (IDS) works as shown in Fig.~\ref{fig:system}. The attackers can access the CAN through various wireless networks. The injected attack actions may control the function of Electronic Control Units (ECUs). The ECUs are the components of the control system of connected vehicles. Therefore, the attacks on ECUs of vehicles through the CAN may raise severe security issues. To mitigate this, we developed a DBL model to detect the attacks. Due to the noise of action or disguise, we cannot eliminate Aleatoric uncertainty (statistical uncertainty). IDS can raise warnings or suspend the abnormal operations based on the confidence level of the predictions to reduce risk. The continuous learning by the model requires the storage of operations and labeling of data by a security engineer in situations where the model has low confidence.



Unlike training neural networks, Bayesian networks weigh uncertainty~\cite{pmlrblundell15,Wilson2020}, which can be given
\begin{equation}
\begin{aligned}
\label{eqn_bayes}
P(w | D) = \frac{P(D | w) P(w)}{\int P(D | w') P(w') \text{d}w'}.
\end{aligned}
\end{equation}
We need to specify the prior density $P(w)$, using training data $D$ to determine the likelihood $P(D | w)$. The normalisation constant is ${\int P(D | w') P(w') \text{d}w'} = P(D)$, which is often intractable; therefore,  a variational posterior $q(w | \theta)$ is often computed to approximate the posterior $P(w | D)$. The difference between them can be computed with Kullback–Leibler divergence, given by 

\begin{equation}
\begin{aligned}
\label{eqn_kl}
D_{KL} (q(w | \theta) || P(w | D)) &= \int q(w | \theta) \log \left( \frac{q(w | \theta)}{P(w | D)} \right) \text{d} w 
\end{aligned}
\end{equation}
To update $\theta$ and minimize the difference between $q(w | \theta)$ and $P(w | D)$, the loss function can be given,

\begin{equation}
\begin{aligned}
\label{eqn_loss}
L(\theta | D) &= D_{KL} ( q(w | \theta) || P(w) ) - \mathbb{E}_{q(w | \theta)}(\log P(D | w)) \\
&= \int q(w | \theta) ( \log q(w | \theta) - \log P(D | w) - \log P(w) ) \text{d}w.
\end{aligned}
\end{equation}
Since the integral over $w$ is computationally expensive, the above function can be re-written as expectation and then use reparameterization~\cite{pmlrblundell15} to update the parameters $\theta$. 
\begin{equation}
\begin{aligned}
\label{eqn_loss1}
L(\theta | D) &= \mathbb{E}_{q(w | \theta)} ( \log q(w | \theta) - \log P(D | w) - \log P(w) ).
\end{aligned}
\end{equation}
The rest of the update process is the same as standard backpropagation~\cite{Hinton1986}, and updates the parameters from the last layer to the first layer.

\section{Results Analysis and Discussion}
\begin{figure}[h!]
  \centering
  \begin{minipage}[b]{0.45\textwidth}
  \includegraphics[width=2.5in]{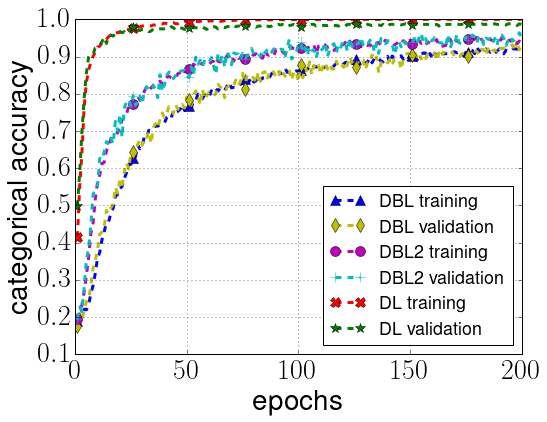}
    \caption{Categorical Accuracy}
    \label{fig:acc}
  \end{minipage}
  \hfill
  \begin{minipage}[b]{0.45\textwidth}
    \includegraphics[width=2.5in]{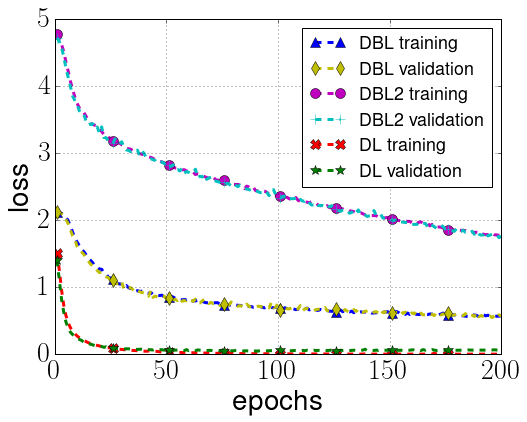}
    \caption{Loss}
    \label{fig:loss}
  \end{minipage}
\end{figure}

\begin{figure}[h!]
  \centering
  \begin{minipage}[b]{0.45\textwidth}
    \includegraphics[width=2.5in]{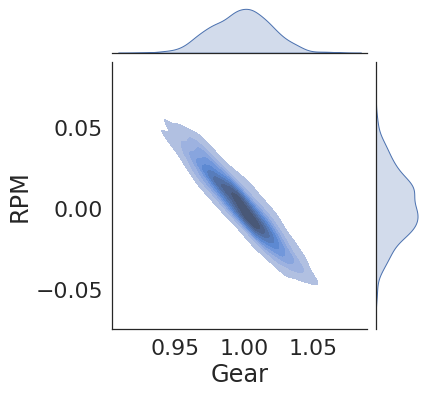}
    \caption{Correlation}
    \label{fig:corr1}
  \end{minipage}
  \hfill
  \begin{minipage}[b]{0.45\textwidth}
  \includegraphics[width=2.5in]{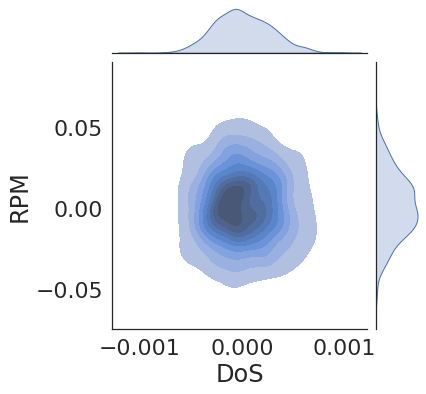}
    \caption{Correlation}
    \label{fig:corr}
  \end{minipage}
\end{figure}

We use a real-world dataset, which can be requested from the website~\cite{song_kim}.
We adopt TensorFlow~\footnote{https://www.tensorflow.org} to build the deterministic deep learning model and TensorFlow probability~\footnote{https://www.tensorflow.org/probability} for the Deep Bayesian Learning (DBL) models. Each operation and attack action has fixed-length binary numbers, and multiple (sequential) actions can be represented as a 2-dimensional feature that can be fed into convolutional neural networks. As we can see from Fig.~\ref{fig:acc} and Fig.~\ref{fig:loss}, the deterministic deep learning model can achieve slightly higher categorical accuracy and less loss than DBL models. However, the deterministic deep learning mode is slightly overfitting after 150 epochs. Furthermore, the deterministic deep learning model cannot provide rich information such as correlation as shown in Fig.~\ref{fig:corr1} and Fig.~\ref{fig:corr} that can assist security engineers to further analyze and identify the risks.

\begin{figure}[h]
    \centering
    \begin{subfigure}[b]{0.4\textwidth}
    \includegraphics[width=\textwidth]{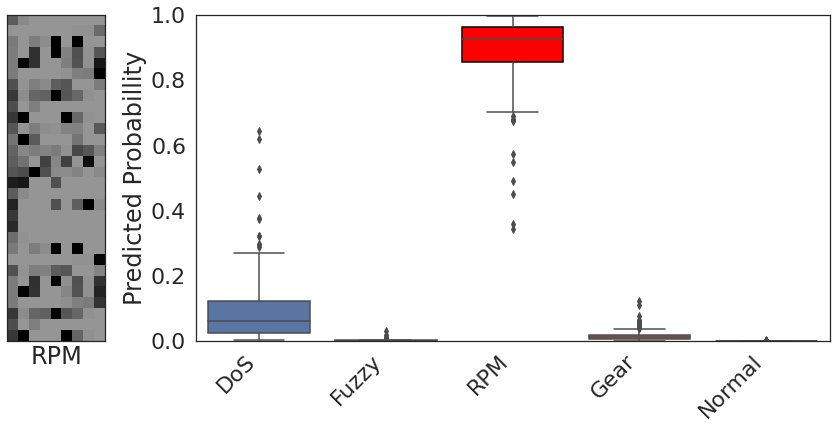}
    \caption{example 1}
    \label{fig:pred1}
    \end{subfigure}
    \hfill
    \begin{subfigure}[b]{0.4\textwidth}
    \includegraphics[width=\textwidth]{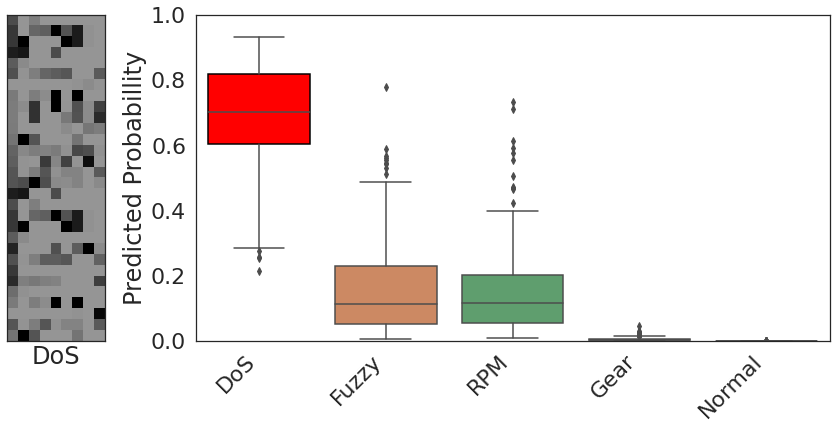}
    \caption{example 2}
    \label{fig:pred2}
    \end{subfigure}
    \hfill
    \begin{subfigure}[b]{0.4\textwidth}
    \includegraphics[width=\textwidth]{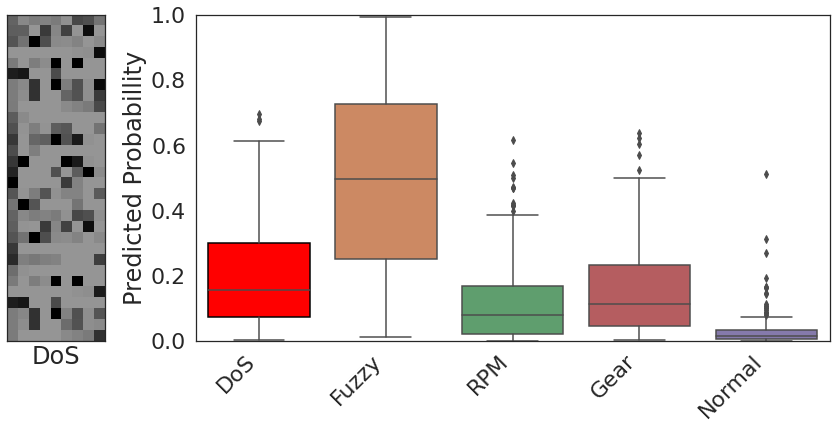}
    \caption{example 3}
    \label{fig:pred3}
    \end{subfigure}
    \hfill
    \begin{subfigure}[b]{0.4\textwidth}
    \includegraphics[width=\textwidth]{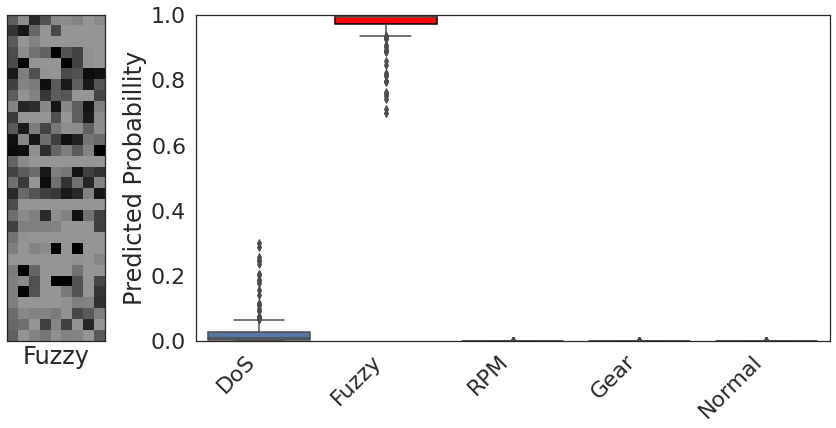}
    \caption{example 4}
    \label{fig:pred3}
    \end{subfigure}
 \caption{Prediction Examples}
 \label{fig:pred}
\end{figure}

Moreover, BDL models can show the uncertainty of its predictions instead of being overconfident like the deterministic models. The models show they do not know when the confidence is low. Our dataset~\cite{Song8514157} has five actions types, including normal, DoS, Fuzzing, RPM, and Gear Spoofing attacks, and the true label is colored red in Fig.~\ref{fig:pred}. Although the model has classified the operations correctly in the example Fig.~\ref{fig:pred}-(a) and (b), it still shows the uncertainty of the predictions, which is given as the probability of each class. In Fig.~\ref{fig:pred}-(c) the model classified the operation incorrectly but was not overly confident about the prediction. In the real world, we can take countermeasures given the confidence of the prediction of BDL models but not with deterministic models. Note that  probability or so called confidence is not actually show the confidence because results are normalized.

Theoretically, we can reduce the Epistemic uncertainty by collecting more data and improving the performance of the models. However, the Bayesian or probabilistic models often require more data to reduce the Epistemic uncertainty than deterministic models. Moreover, it is challenging
to set up prior distributions and incorporate prior knowledge into complexity models like Bayesian networks.

\section{Conclusions}
In this work, we proposed an Intrusion Detection System using a Deep Bayesian Learning (DBL) model. We compare the DBL model with deterministic deep learning by using both types of models on a real-world dataset and show the advantages of the DBL model. Specifically, the DBL model can provide more information about its prediction than deterministic models. The additional information can help security engineers with further analysis of abnormal behaviours and label the data. The DBL model can provide the uncertainty of the prediction and show its confidence of the predictions. However, the DBL model has a higher Epistemic uncertainty and is difficult to incorporate prior knowledge and reduce training cost. For future work, we will investigate training methods that can reduce Epistemic uncertainty. 

\nocite{*}
\bibliographystyle{IEEEannot}
\bibliography{annot}
\section{Appendix}

\begin{figure}[h]
    \centering
    \begin{subfigure}[b]{0.4\textwidth}
    \includegraphics[width=\textwidth]{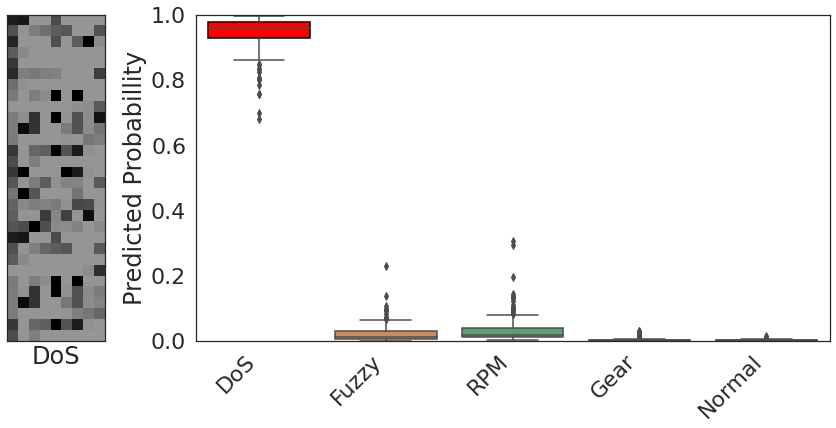}
    \caption{example 1}
    \label{fig:pred1}
    \end{subfigure}
    \hfill
    \begin{subfigure}[b]{0.4\textwidth}
    \includegraphics[width=\textwidth]{images/pred_xmp2.png}
    \caption{example 2}
    \label{fig:pred2}
    \end{subfigure}
    \hfill
    \begin{subfigure}[b]{0.4\textwidth}
    \includegraphics[width=\textwidth]{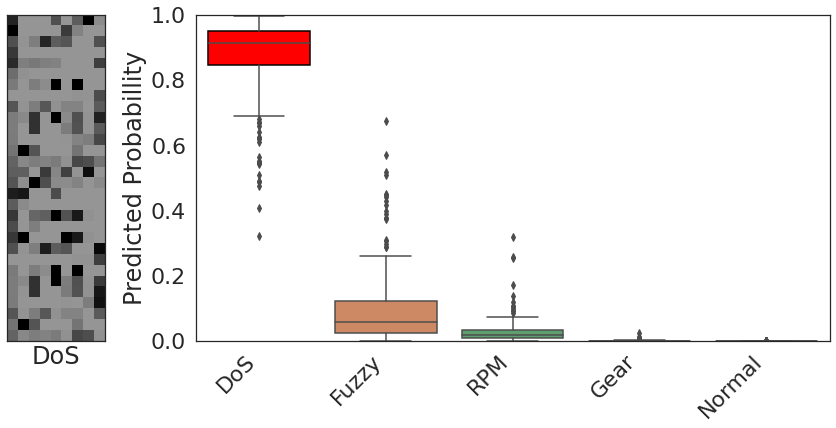}
    \caption{example 3}
    \label{fig:pred3}
    \end{subfigure}
     \hfill
    \begin{subfigure}[b]{0.4\textwidth}
    \includegraphics[width=\textwidth]{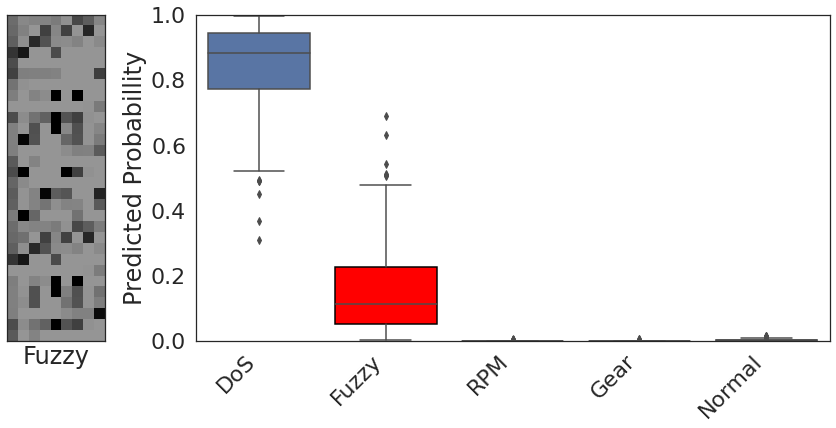}
    \caption{example 4}
    \label{fig:pred1}
    \end{subfigure}
    \hfill
    \begin{subfigure}[b]{0.4\textwidth}
    \includegraphics[width=\textwidth]{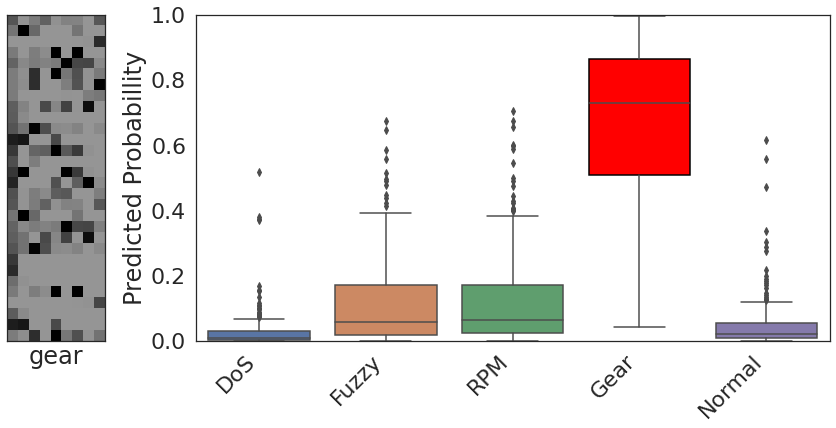}
    \caption{example 5}
    \label{fig:pred2}
    \end{subfigure}
    \hfill
    \begin{subfigure}[b]{0.4\textwidth}
    \includegraphics[width=\textwidth]{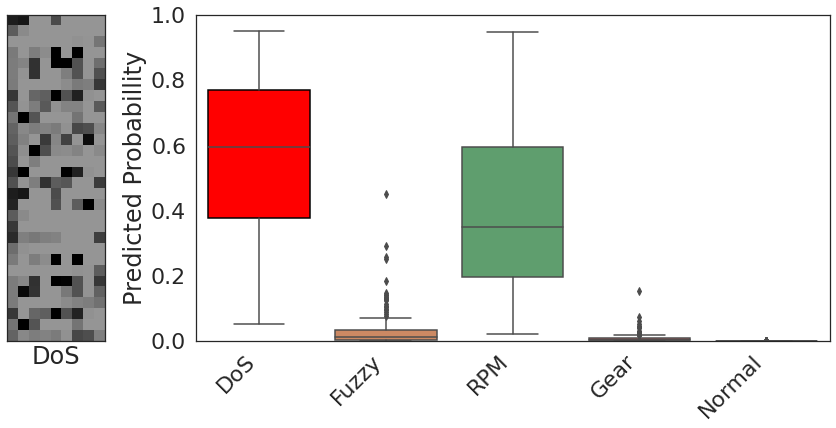}
    \caption{example 6}
    \label{fig:pred3}
    \end{subfigure}

 \caption{More Prediction Examples.}
 \label{fig:pred2}
\end{figure}
Note that most of the predictions have high confidence. The prediction examples show the uncertainty of the predictions.

\end{document}